\documentclass[final,1p,times]{elsarticle} 
\usepackage{graphicx} 
\usepackage{amssymb} 
\usepackage{amsthm} 
\usepackage{lineno} 

\def\nc{ {N_{c}}}
\def\nf{ {N_{f}}}
\def\vl{ {v_{\rm lim}}}
\def\zm{{z_{\rm m}}}
\def\wp{ {\omega_{\rm peak}}}

\newcommand{\jpsi}{J/\psi}

\def\be{\begin{eqnarray}}
\def\ee{\end{eqnarray}}

\def\Fig#1{Fig.~\ref{#1}}
\def\Eq#1{Eq.~(\ref{#1})}

\journal{Nuclear Physics A} 

\begin{document} 


\begin{frontmatter} 


\title{A Photon Peak due to Strong Coupling}

\author{Jorge Casalderrey-Solana}

\address{Physics Department, Theory Unit, CERN, CH-1211 Gen\'eve, Switzerland}

\begin{abstract} 
We show that if a flavour-less vector meson remains bound after deconfinement, and if its limiting velocity in the quark-gluon plasma is subluminal, then this meson produces a distinct peak in the spectrum of thermal photons emitted by the plasma. We also demonstrate that this effect is a universal property of all strongly coupled, large-$\nc$ plasmas with a gravity dual. For the $J/\Psi$ the corresponding peak lies between 3 and 5 GeV and could be observed  at the LHC. 

\end{abstract} 

\end{frontmatter} 



\section{Introduction}\label{}
In recent years several experimental and theoretical considerations indicate that the QGP right above the deconfining phase transition 
does not behave like a weakly coupled gas of quarks and gluons but that the interactions among plasma constituents are not 
small. This observation has lead to increasing interest in the dynamics of strongly coupled plasmas analyzed via the 
AdS/CFT correspondence. In these proceedings we report about a generic feature of all strongly coupled plasmas with a gravity
dual \cite{CasalderreySolana:2008ne}. 

Our discussion is based  on two assumptions about the properties of heavy mesons in a deconfined medium which are
fulfilled  by all gauge theories with a gravity dual:
\begin{enumerate}
\item Heavy vector mesons such as the $\jpsi$ remain bound above the deconfinement transition. Even though this is the
subject of intense theoretical research, studies of lattice data indicate that this might be the case \cite{jpsiinmed}
\item The in-medium dispersion relation of these mesons is significantly modified and they posses a (sub-luminal) maximum speed of propagation.  We will also need to assume that the dispersion relation extends to sufficiently large momentum
and that the meson states remain sufficiently  narrow.
\end{enumerate}

\begin{figure}
\includegraphics[scale=.35]{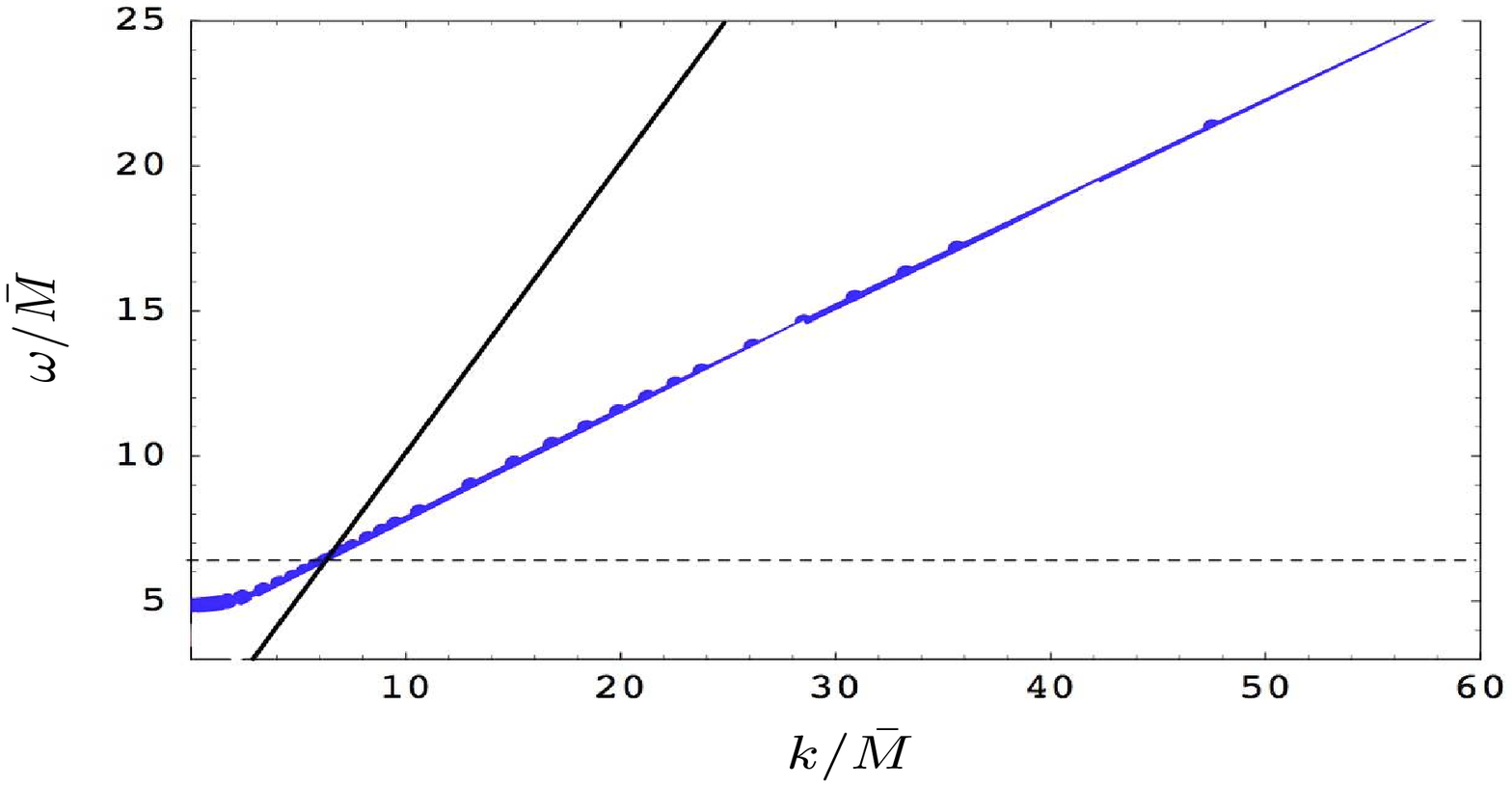}
\includegraphics[scale=.40]{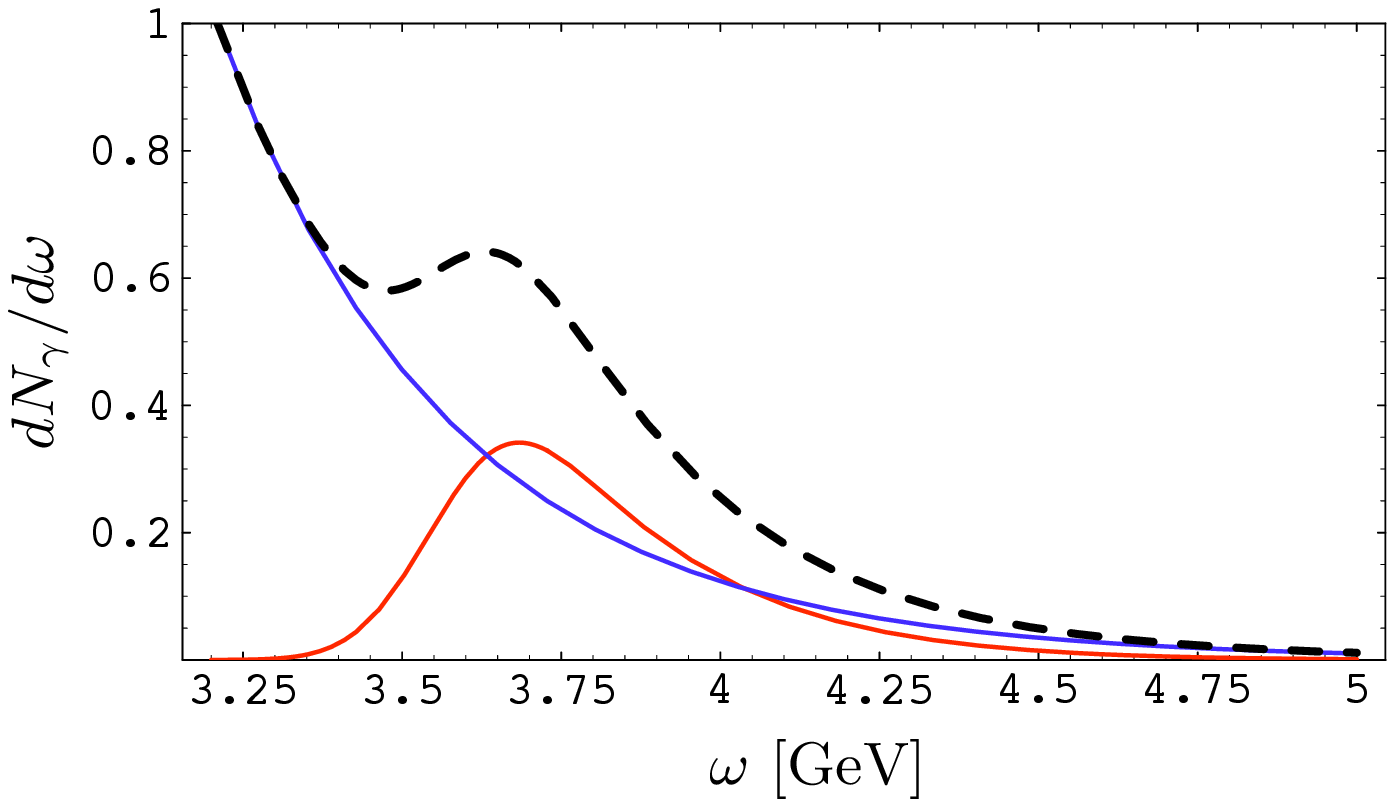}
\caption{Left:Dispersion relation (blue curve) for a heavy meson in the ${\cal N}=4$ SYM plasma at strong coupling \cite{MMT2}. The black straight line corresponds to 
$\omega=k$.
Right:
Thermal photon spectrum for LHC energies with arbitrary normalization. The continuous, monotonically decreasing, blue curve is the background from light quarks. The continuous, red curve is the signal from $\jpsi$ mesons. The dashed, black curve is the sum of the two.\label{thefigures}}
\end{figure}

These assumptions are enough to describe the main features of the in-medium dispersion relation; its generic form 
is shown in \Fig{thefigures} a). At high momentum, the dispersion relation approaches a constant slope, which determines
the in-medium limiting velocity $\vl$. At small momentum the dispersion relation approaches a finite value of $\omega$, the in medium mass.  As a consequence of this asymptotic behavior, the dispersion relation crosses the light-cone 
at a position $k=\omega=\wp$. Note that this crossing does not mean that heavy quark becomes massless at this point;
the dispersion relation is very different from the vacuum one and there is no reason why the invariant $\omega^2-k^2$ 
should be considered as the mass. 

At the point $\wp$ the in-medium $\jpsi$ has the same quantum numbers of the photon; thus, the electromagnetic
interaction leads to mixing of these two degenerate states. What this means is that at $\wp$ the in-medium
$\jpsi$ can decay into an {\it on shell} photon.  This is different from the standard $\jpsi$ decay into dileptons, in which
the $\jpsi$ leads to an {\it off shell} photon of virtuality of the $\jpsi$ mass. The mixing between the in-medium 
$\jpsi$ and the on shell photon leads to a peak in the  current-curren correlator at null momentum 
$
\chi(\omega,T)=\left<
			J^\mu(\omega, k=\omega) J_\mu(-\omega, k=-\omega)
			\right> 
$
at $\omega=\wp$.

A direct consequence of this observation is  that if the dispersion relation of mesons is of the type described, a quark 
gluon plasma at finite temperature with a  $\jpsi$ in kinetic equilibrium would lead to an almost monochromatic peak of 
photons  at $\wp$ with a width given by the in-medium width of the $\jpsi$.

\section{A universal property of plasmas with a gravity dual}
We can find an example in which the assumed features of the meson dispersion relation are realized: $\mathcal{N}=4$
super-symmetric Yang Mills with $\nf$ flavors  in the  $\nc\rightarrow\infty$, $\lambda\rightarrow\infty$ limit with 
$\nf/\nc\rightarrow 0$ at finite temperature. In this case, the dynamics can be  described by a dual gravitational
theory with $N_f$ D7 probe branes; the fact that $\nf/\nc\rightarrow 0$ implies that the back-reaction of the branes 
can be neglected and the metric  is that of thermal $\mathcal{N}=4$ without D7 branes which in suitable coordinates
is given by 
\be
ds^2=\frac{L^2}{z^2} \left(-f(z) dt^2 + \frac{dz^2}{f((z)} + dx^2\right) + L^2 d\Omega_5
\ee
with $f=1-\left(z \pi T\right)^4$ and T is the temperature. The D7 branes span the coordinates $t,\,x,\,z$ and a 3-sphere
within the $\Omega_5$. The D7 branes can end "in thin air" at a scale $\zm$; if $\zm<1/T$ the branes are outside the
black  hole (Minkowski embedding). In this case, the small excitations of the brane correspond to (infinitely narrow) mesons in the dual gauge theory and the 
scale $1/\zm$ determines their mass. If the temperature is higher, the brane falls inside of the black-hole and there are
no mesons excitations (black-hole embedding) \cite{MMT}.  The D3-D7 model in the Minkowski phase constitutes an 
interesting model to study the dynamics of meson in a strongly coupled, deconfined gauge theory  plasma.

In the Minkowski embedding, due to the gravitational attraction, the meson wave function is  concentrated around
the tip of the brane, $\zm$. As a consequence of the gravitational redshift in the presence of the black hole, the 
speed of light in the $\left(t,x\right)$ hyperplane is reduced and it is given by
\be
\label{eqvl}
\vl=\sqrt{-\frac{g_{tt}}{g_{xx}}}=1-\left(\pi T \zm\right)^4
\ee
 thus, meson excitations in this theory cannot move faster than $\vl$. This observation provides a simple geometrical
interpretation for  the meson 
dispersion relation computed in the D3/D7 model \Fig{thefigures} a).

We now note that discussion above is common to all  strongly coupled, large $\nc$ gauge theory plasmas with a gravity 
dual since in these theories a finite number of flavors $\nf$ is described by $\nf$ D-brane probes and since the deconfined
phase of these theories is described by background with a black-hole. Since at the black-hole horizon the 
time component of the metric $g_{tt}\rightarrow0$ the limiting velocity of the mesons excitations \Eq{eqvl} is sub-luminal.
Thus, if QCD has a gravity dual in the large $\nc$ limit, the dispersion relation of heavy mesons
(and in particular heavy vector mesons) must cross the lightcone leading to  a peak on the photon spectrum as discussed in the previous section.  

\section{A photon peak at the LHC}
We now address whether such a peak could be observed in a relativistic nucleus-nucleus collision. 
The natural vector meson candidates to consider are the $\jpsi$ and the $\Upsilon$ since they may survive the 
deconfining transition (we have  not considered the $\rho$ since its survival is more controversial). The main difference
with the analysis above is that in a collision the temperature of the plasma is not fixed, but it evolves as the 
plasma expands and cools. As a consequence, the energy at which the peak would occur changes: it grows as the
temperature is reduced. Depending on how fast the expansion happens, the peak may be transformed into an 
enhancement in the neighborhood  of the meson mass. Another difficulty for the observation of the peak is the 
existence of more sources for photons such as the thermal emission from light degrees of freedom and the 
prompt photon emission (perturbative contribution).

Addressing the exact magnitude of the effect in a nuclear collision is a complicated task that demands an accurate model
of the dynamics of the medium, the dynamics of mesons in the plasma and the meson production mechanism. However
there are significant theoretical uncertainties in these issues, specially in the interaction of meson with the plasmas; there 
are different theoretical models that address the $\jpsi$ suppression pattern. Thus, we will not try to improve the current
description of in-medium meson, but we will construct a model in which the peak can be observed in the final 
photon distribution. 

Instead of embedding the thermal rate into a hydrodynamical calculation, for this  estimate we use a fireball model
in which the volume of the system expands as $V(t)=\pi(z_0+v_z \, t)(r_0 + a_\perp \, t^2/2)^2$ and, in the QGP phase
the product $V(t) T(t)^3$ is constant. 
We will assume that the initial temperature of the fireball is larger than the meson dissociation temperature. 

We first consider the $J/\psi$ since the thermal photon emission in the region of $\omega=3-5$ GeV is dominant at the LHC. Inspired by statistical recombination 
\cite{statistical}, we assume that the $c\bar{c}$-pairs produced in the initial  collisions become kinetically equilibrated in the QGP 
 but their total number, $N_{\rm c\bar{c}}$, stays constant. This condition is implemented via a fugacity factor 
\be
g_{\rm c}(T) = \frac{N_{\rm c\bar c}}{2 \cdot 3 \cdot V(T) \, 
\left( \frac{M_{\rm c} T}{2 \pi} \right)^{3/2} e^{-M_{\rm c}/T}} \,,
\ee
where $M_{\rm c}$ is the in-medium charm-quark mass. At $t=t_{\rm diss}$ the temperature reaches the dissociation temperature of the $J/\psi$ in the medium and a fraction of the $c\bar{c}$-pairs recombine forming $J/\psi$ mesons. Their contribution to the total number of photons is
\be
S(\omega) \propto \int_{t_{\rm diss}}^{t_{\rm hadro}} dt \, 
V(t) \, g_{\rm c}(T(t))^2 \,  
e^{-\omega/T(t)}
\, \chi_{{\rm J/\psi}} 
(\omega,T(t)) \,,
\label{signal}
\ee
where the `S' stands for `signal'. The overall normalization is not important since it is the same as that for the background of thermal photons emitted by the light quarks, which is given by
\be
B(\omega) \propto \int_{0}^{t_{\rm hadro}} dt \, V(t) \, e^{-\omega/T(t)}  \,.
\label{B}
\ee
We have omitted $\chi_{{\rm light}}(\omega,T)$ since, guided by the results for plasmas with a gravity dual, we assume it is structureless. 

We  model the spectral function for the $\jpsi$ by a unit-area Gaussian distribution of width $\Gamma=100$ MeV  centered at  
$\wp(T(t))$. The dispersion relation we use is a fit to \Fig{thefigures} a) and the $N_{\rm c\bar c}=60$. The $\jpsi$
dissociation temperature is $T_{diss}=1.25 \, T_{\rm c}$.
The results of numerically evaluating \Eq{signal} and \Eq{B} for LHC values of the parameters are plotted in 
\Fig{thefigures} b) where the effect of the $\jpsi$ peak is clearly seen on the spectrum. However, this figure should
be taken as illustrative, since the precise form is very sensitive to the  value of the parameter used. The enhancement is, in particular, is quadratically sensitive to $N_{\rm c\bar c}=60$; thus, our estimates indicate that at 
RHIC, where $N_{\rm c\bar c}$ is ten times smaller, the enhancement cannot be observed. We have also explored 
the possibility of a similar enhancement for the
$\Upsilon$ meson at $\wp \gtrsim 10$ GeV; however at these energies the thermal photon contribution is exponentially
suppressed as compare to pQCD photons.

As a conclusion, we have seen that a modified dispersion relation of the type shown in \Fig{thefigures} a) for the $\jpsi$ leads, under certain assumptions, to a distinct peak in the thermal photon yield at 
the LHC in the region of $3-5$ GeV. This dispersion relation is expected in all strongly coupled gauge theory plasma, so its observation would be in agreement with the interpretation of the QGP as strongly coupled. Unfortunately, there are  many model assumptions in the analysis and not observing this peak 
would not necessarily imply a weakly coupled picture of the QGP.

\section*{Acknowledgments} 
This work was done in collaboration with D. Mateos and it has been supported by a Marie Curie Intra-European Fellowship of the European Community's Seventh Framework Programme under contract number (PIEF-GA-2008-220207)

\end{document}